\documentclass[aps,pra,twocolumn,showpacs]{revtex4}
\usepackage{graphicx}
\usepackage{amsmath,amsfonts}

\newcommand{\er}{\mathbf{r}}
\newcommand{\ee}{{\rm e}}

\begin{document}

\title{Controlled spin domain creation by phase separation}

\author{Tomasz \'Swis\l{}ocki}
\affiliation{Instytut Fizyki PAN, Aleja Lotnik\'ow 32/46, 02-668 Warsaw, Poland}

\author{Micha\l{} Matuszewski}
\affiliation{Instytut Fizyki PAN, Aleja Lotnik\'ow 32/46, 02-668 Warsaw, Poland}

\begin{abstract}
We demonstrate a method of controlled creation of spin domains in spin-1 antiferromagnetic Bose-Einstein condensates.
The method exploits the phenomenon of phase separation of spin components in an external potential.
By using an appropriate time dependent potential, a composition of spin domains can be created, as
we demonstrate in the particular cases of a double well and a periodic potential.
In contrast to other methods, which rely on spatially inhomogeneous magnetic fields, here the domain structure is completely
determined by the optical fields, which makes the method versatile and reconfigurable. It allows
for creation of domains of various sizes, with the spatial resolution limited by the spin healing length only.
\end{abstract}

\maketitle

\section{Introduction}

Bose-Einstein condensates with spin degrees of freedom~\cite{Ho} attracted in recent years 
great interest due to the unique possibility of exploring fundamental concepts of quantum mechanics
in a remarkably controllable and tunable environment.
The ability to generate spin squeezing and entanglement~\cite{Entanglement}
makes spinor Bose gases promising candidates for 
applications as quantum simulators~\cite{QS}, in quantum information~\cite{QI}, and for precise measurements~\cite{Measurement}.
Moreover, spinor condensates were successfully used to recreate many of the phenomena of condensed matter physics
in experiments displaying an unprecedented level of control over the quantum system.
In particular, spin domains~\cite{Stenger_Nat_1998,Quenched,Ketterle_Metastable}, 
spin mixing~\cite{Mixing}, and spin vortices~\cite{Ketterle_Coreless}
were predicted and observed.

The ability to create spin domains is a crucial component of applications including data storage and 
spin based logic implementation~\cite{DomainWallsApps}. To date, domain structures
in Bose-Einstein condensates were created in a controllable way using inhomogeneous magnetic fields~\cite{Stenger_Nat_1998}.
In principle, these can be induced by magnetic coils, 
electronic chips or permanent magnets~\cite{MagneticDomainCreation}. However, both the spatial resolution 
of this method and the ability
to produce arbitrarily shaped, reconfigurable spin domain structures is severely limited.

In recent papers~\cite{Matuszewski_PS,Matuszewski_Trap}, we considered the possibility of phase separation
in the ground state of spinor condensates. We demonstrated that this phenomenon can take place in both
antiferromagnetic and ferromagnetic condensates in external potentials.
In this paper, we exploit the phenomenon of phase separation to
propose a method of controlled creation of metastable spin domain states with a chosen spatial spin structure.
It consists of applying an appropriately designed optical potential, 
which is subsequently slowly relaxed towards the desired final trapping potential.
In contrast to the other method, relying on magnetic field gradients to separate spin components,
our method uses more easily reconfigurable optical laser fields to shape the spatial structure.
Importantly, these structures are not ``pinned'' by local extrema of the magnetic field.
An additional advantage is the high spatial resolution, which we demonstrate to be generally
limited only by the spin healing length of the condensate. 

The paper is organized as follows. Section~\ref{Sec_model} reviews the mean-field model of a spin-1 condensate 
in a homogeneous magnetic field. In Section~\ref{Sec_numerical} we present numerical simulations
of the spin domain creation process. In Section~\ref{Sec_limitations} we discuss practical limitations of the proposed method
using both analytical and numerical approach. Section~\ref{Sec_conclusions} concludes the paper.

\section{Model} \label{Sec_model}

We consider a spin-1 Bose-Einstein condensate in a homogeneous magnetic field pointing along the $z$ axis.
We apply the mean field approximation, which describes dilute Bose-Einstein condensates
at zero temperature. In spinor condensates, ground states can substantially deviate 
from the mean field solutions even at zero temperature due to large quantum fluctuations~\cite{NonMeanField}.
However, it was shown that the introduction of a weak magnetic field restores the validity of the mean-field model.
We start with the Hamiltonian $H = H_{\rm S} + H_{\rm A}$,
\begin{equation} \label{En}
H = \sum_{j=-,0,+} \int d\er \, \psi_j^* \left(-\frac{\hbar^2}{2m}\nabla^{2} + \frac{c_0}{2} n 
+ V({\bf r})\right) \psi_j + H_A,
\end{equation}
where $\psi_-,\psi_0,\psi_+$ are the wave functions of atoms in magnetic sublevels $m_{\rm f}=-1,0,+1$, 
$m$ is the atomic mass, $V({\bf r})$ 
is an external potential and $n=\sum n_j = \sum |\psi_j|^2$ is the total atom density.  
The asymmetric (spin dependent) part of the Hamiltonian is given by
\begin{equation}
\label{EA}
H_{\rm A} = \int d\er \left(\sum_{j=-,0,+} E_jn_j + \frac{c_2}{2}|{\bf F}|^2\right)
\end{equation}
where $E_j$ is the Zeeman energy shift for state $\psi_j$ and the spin density is,
\begin{equation}
\label{spindensity}
{\bf F}=(F_x,F_y,F_z)=(\psi^{\dagger}\hat{F}_x\psi,\psi^{\dagger}\hat{F}_y\psi,\psi^{\dagger}\hat{F}_z\psi)
\end{equation}
where $\hat{F}_{x,y,z}$ are the spin-1 matrices~\cite{Isoshima_PRA_1999} and $\psi =(\psi_+,\psi_0,\psi_-)$.  
The spin-independent and spin-dependent interaction coefficients are given by $c_0=4
\pi \hbar^2(2 a_2 + a_0)/3m$ and $c_2=4 \pi \hbar^2(a_2 -
a_0)/3m$, where $a_S$ is the s-wave scattering length for colliding atoms
with total spin $S$.
The total number of atoms and the total magnetization in the direction of the magnetic field
\begin{align}
N&=\int n d \er\,, \\
M&= \int F_z d \er = \int \left(n_+ -
n_-\right) d \er\,,
\end{align}
are conserved quantities.
The Zeeman energy shift for each of the components, $E_j$ can be calculated using the
Breit-Rabi formula~\cite{Wuster}
\begin{align}
E_{\pm}& = -\frac{1}{8}E_{\rm HFS}\left(1 + 4\sqrt{1\mp \alpha + \alpha^2} \right)  \mp g_I \mu_B B\,, \nonumber \\
E_{0} &= -\frac{1}{8}E_{\rm HFS}\left(1 + 4\sqrt{1 + \alpha^2} \right)\,,
\label{BR}
\end{align}
where $E_{\rm HFS}$ is the hyperfine energy splitting at zero
magnetic field, $\alpha = (g_J - g_I) \mu_B B/E_{\rm HFS}$, where
$\mu_B$ is the Bohr magneton, $g_I$ and $g_J$ are the nuclear and electronic g-factors, 
and $B$ is the magnetic field strength.
The linear part of the Zeeman effect gives rise to an
overall shift of the energy, and so we can remove it with the
transformation
\begin{equation}
H \rightarrow H + (N + M) E_+/2 + (N - M) E_-/2\,.
\end{equation}
This transformation is equivalent to the removal of the Larmor precession
of the spin vector around the $z$ axis \cite{Matuszewski_PRA_2008}.
We thus consider only the effects of the quadratic Zeeman shift.
For sufficiently weak magnetic field we can approximate it by $\delta
E=(E_+ + E_- - 2E_0)/2  \approx \alpha^2 E_{\rm HFS}/16$, which is 
positive for $^{87}$Rb and $^{23}$Na condensates.

The Hamiltonian~(\ref{En}) gives rise to the Gross-Pitaevskii equations describing the mean-field dynamics of the system
\begin{align}\label{GP}
i \hbar\frac{\partial \psi_{\pm}}{\partial t}&=\left[ \mathcal{L} +
c_2 (n_{\pm} + n_0 - n_{\mp})\right] \psi_{\pm} +
c_2 \psi_0^2 \psi_{\mp}^* \,, \\\nonumber
i \hbar\frac{\partial \psi_{0}}{\partial t}&=\left[ \mathcal{L} -
\delta E + c_2 (n_{+} + n_-)\right] \psi_{0} + 2 c_2
\psi_+ \psi_- \psi_{0}^* \,,
\end{align}
where $\mathcal{L}$ is given by $\mathcal{L}=-\hbar^2\nabla^2/2m+c_0n + V({\bf r})$.

By comparing the kinetic energy with the interaction energy, we can determine the
healing length $\xi=2\pi\hbar / \sqrt{2m c_0 n}$ and the spin healing length
$\xi_s=2\pi\hbar / \sqrt{2m c_2 n}$. These quantities give the length scales of
spatial variations in the condensate profile induced by the spin-independent or spin-dependent
interactions, respectively. Analogously, we define the magnetic healing length as
$\xi_B=2\pi\hbar / \sqrt{2m \delta E}$.

In spin-1 condensates created to date, the $a_0$ and $a_2$ scattering lengths have similar magnitudes.
The spin-dependent interaction coefficient $c_2$ is then much smaller than its spin-independent
counterpart $c_0$. For example, this ratio is about 1:30 in a $^{23}$Na condensate and 1:220 in a $^{87}$Rb condensate 
far from Feshbach resonances \cite{Beata}.
Consequently, changing the total density $n$ requires much more energy than changing the relative populations of spin states $n_j$.
In our considerations we will assume that the total atom density profile $n(\er,t)$ is close to the Thomas-Fermi profile
for a given potential $V(\er,t)$.

\section{Spin domain creation} \label{Sec_numerical}

\begin{figure}
\includegraphics[width=9.5cm]{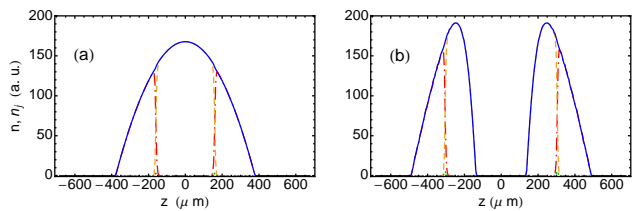}
\caption{(a) Ground state of a sodium condensate with fixed magnetization $M=0.4$, number of atoms $N=8.4\times10^3$, trap frequencies
$\omega_\rho=2\pi\times 1000\,$Hz  and $\omega_z=2 \pi \times 2.5\,$Hz,
and the magnetic field strength $B=120\,$mG. The $n_{+}$, $n_{0}$, and $n_{−}$ components are depicted by dash-dotted, dashed, and dotted lines, respectively,
and the solid lines show the total density. (b) The result of a slow introduction of a Gaussian shaped barrier with $A=3.3\times10^{-30}\,$J and $w=90 \mu$m during the time $t=5\,$s.
}
\label{fig_switchon}
\end{figure}

In Refs.~\cite{Matuszewski_PS,Matuszewski_Trap} we demonstrated that in the presence of a homogeneous magnetic field, 
a spin-imbalanced (magnetized) antiferromagnetic condensate is subject 
to phase separation in the ground state resulting in separate spin domains. 
The spin domains align according to a simple rule; the magnetized domains reside in the regions
with the lowest density, while the unmagnetized domains fill the areas of high density. In this way,
the nonlinear spin energy of the antiferromagnetic condensate is minimized for a given total magnetization. 
In this work, we will show that one can use this property to shape the spatial
distribution of spin with an appropriate time varying external potential.

\begin{figure}
\includegraphics[width=14cm]{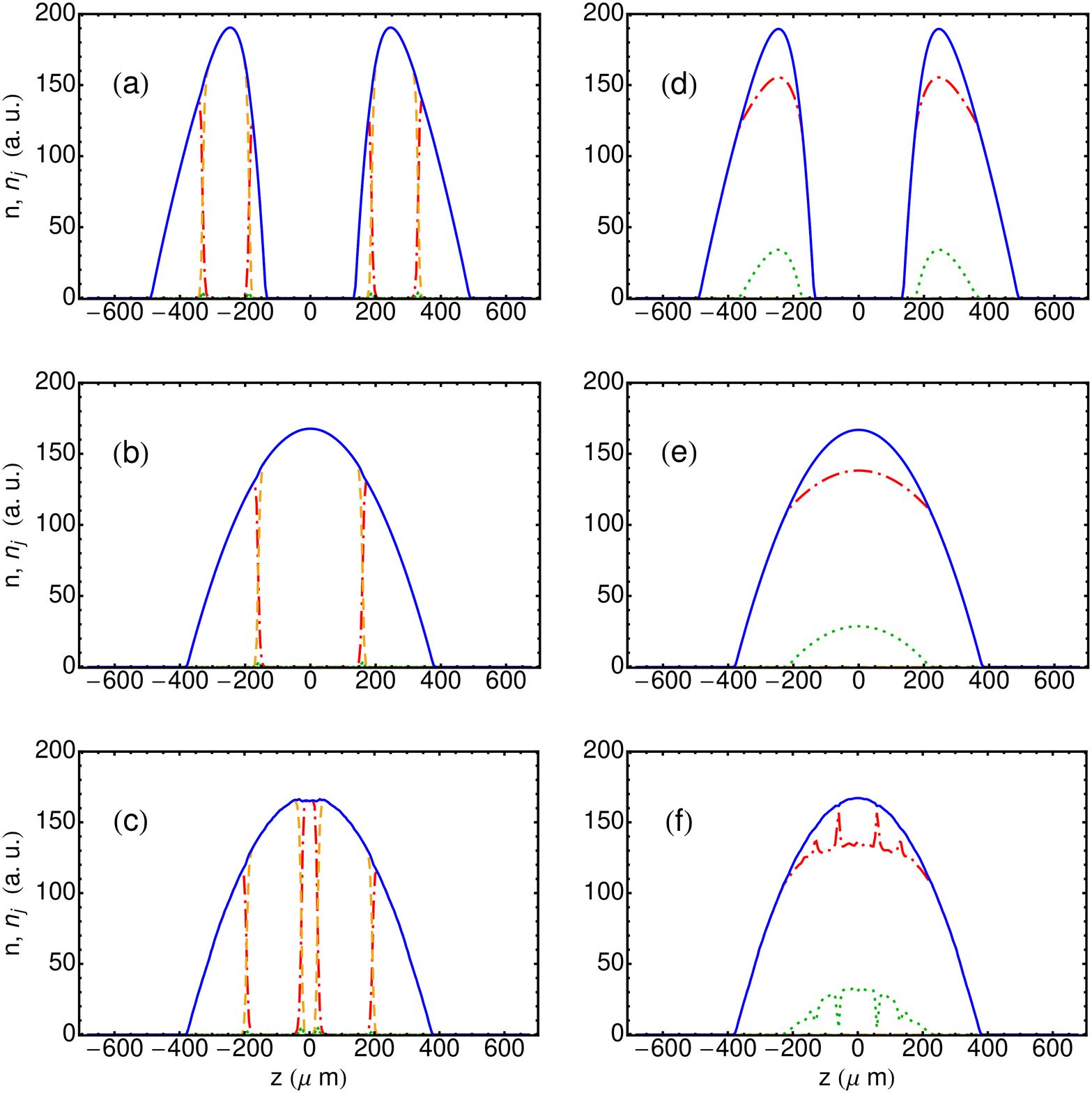}
\caption{
(a) Ground state of the double well potential with parameters as in Fig.~\ref{fig_switchon}. 
(b) Ground state after the removal of the Gaussian barrier. (c) Spin domain structure after the removal of the Gaussian
barrier within $t=5\,$s. (d)-(f) Same as (a)-(c), but for a weaker magnetic field $B=45\,$mG and magnetization $M=0.8$.
}
\label{fig_switchoff}
\end{figure}

We consider a quasi-1D sodium condensate trapped in an elongated harmonic potential
described by the 1D version of Eqs.~(\ref{GP}) \cite{Matuszewski_PS}
with rescaled interaction coefficients $c_0^{\rm 1D},c_2^{\rm 1D} = (m \omega_\perp) / (2\pi \hbar) c_0,c_2$, 
where $\omega_\perp$ is the transverse trapping frequency.
The Fermi radius of the transverse trapping potential is
smaller than the spin healing length, 
and the nonlinear energy scale is much smaller than the transverse 
trap energy scale, which allows us to reduce the
problem to one spatial dimension \cite{Beata,NPSE}. 
To create a double well potential, we add a Gaussian potential barrier that can be realized eg.~by an additional blue-detuned laser beam
\begin{equation}
V(z,t) = \frac{1}{2} m \omega_z^2 z^2 + A(t) \ee^{-z^2/2 w^2}\,.
\end{equation}
In the absence of the barrier, and under certain experimental conditions~\cite{Matuszewski_PS}, the ground state  
is characterized by spatial separation of the $m_f=0$ and $m_f=+1$ atoms 
(or $m_f=-1$, depending on the magnetization), see Fig.~\ref{fig_switchon}(a).
After switching the barrier on, the condensate splits in an effective longitudinal double well potential into two symmetric parts.
We note that the ground state now has the structure corresponding to Fig.~\ref{fig_switchoff}(a). 
However, the result of a slow increase of the barrier on the timescale scale of several seconds 
results in a different state, 
where inner $m_f=+1$ domains are absent on both sides of the barrier. The difference between Fig.~\ref{fig_switchon}(b)
and Fig.~\ref{fig_switchoff}(a) is a result of the strong spin immiscibility~\cite{Ketterle_Metastable} of the $m_f=0,+1$ atoms 
that suppresses the tunnelling of $m_f=+1$ atoms through the $m_f=0$ atoms. The final state is thus a metastable state, 
but practically stable on the relevant experimental timescales.

A yet more intriguing effect can occur when the condensate is prepared in the ground state of the double well potential 
(with the barrier on), Fig.~\ref{fig_switchoff}(a), and after gradually lowering the potential barrier. The condensate, instead of 
evolving towards the single-well ground state, Fig.~\ref{fig_switchoff}(b), creates another kind of a metastable state, with five alternating
spin domains of $m_f=0$ and $m_f=+1$ atoms shown in Fig.~\ref{fig_switchoff}(c). Again, we checked that these spin domains are perfectly stable 
on the timescales as long as several seconds due to suppressed tunneling. In principle, by applying several potential barriers  
with certain parameters, it is possible to create a given number of stable spin domains of various sizes.

We note that the above spin-domain creation scenario is possible only for domains which consist of immiscible atoms. The phase separation
of $m_f=0$ and $m_f=\pm1$ atoms is a feature of antiferromagnetic spin-1 condensates in a relatively strong
magnetic field~\cite{Matuszewski_PS}. In the weaker field regime, the spin-imbalanced
antiferromagnetic condensate generally consists of $m_f=+1$ and $m_f=-1$ atoms, which are miscible. In result, the spin domains
do not separate, and the slow process always results in a state close to 
the ground state of the system, as shown in Fig.~\ref{fig_switchoff}(d)-(f).

\begin{figure}
\includegraphics[width=9.5cm]{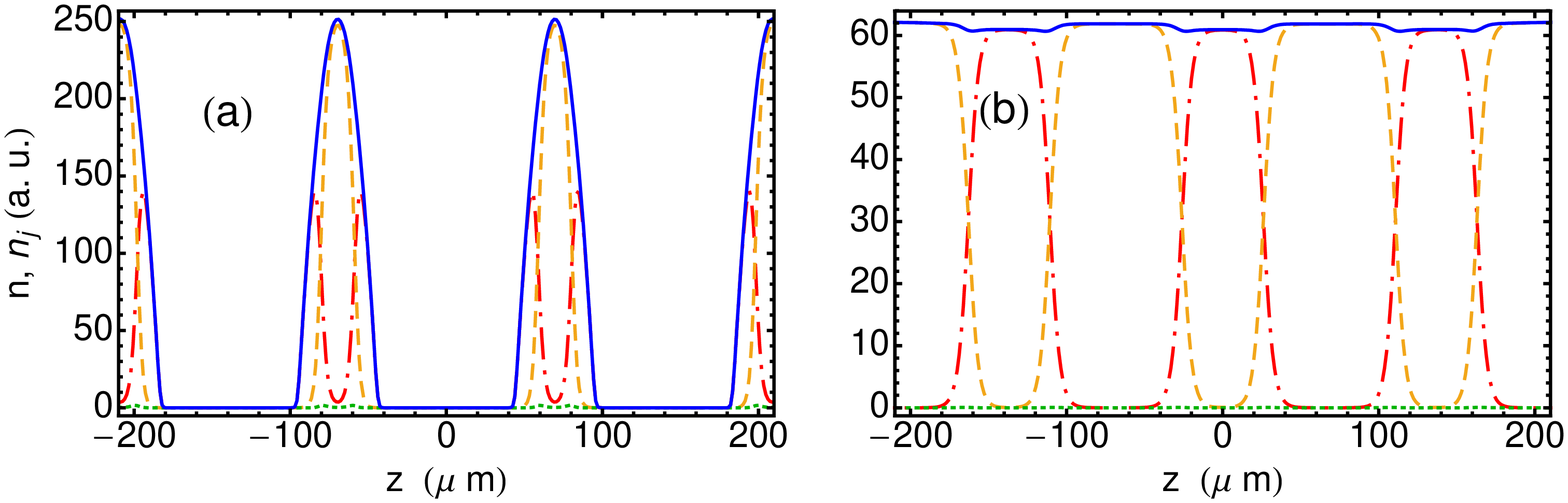}
\caption{
(a) Ground state in a periodic potential with $d=140\,\mu$m and $A=3.3\times10^{-30}\,$J. (b) The resulting domain structure after
switching the potential off during $t=5\,$s.
}
\label{fig_periodic}
\end{figure}
Furthermore, we consider a condensate placed in a periodic optical potential 
\begin{equation}
V(z) = A(t) \cos^2 (2\pi z/d)\,,
\end{equation}
with $d$ being the lattice period. We simulate the dynamics of the condensate with periodic boundary conditions, which
can be experimentally realized eg.~with a ring-shaped trapping geometry~\cite{Toroidal}. Analogously as in the previous case,
we start with a condensate with the optical lattice switched on. the density and spin pattern is shown in Fig.~\ref{fig_periodic}(a).
By gradually decreasing the optical lattice strength, we arrive at a metastable state depicted in Fig.~\ref{fig_periodic}(b),
composed of multiple $m_f=0$ and $m_f=+1$ spin domains. This state is also stable on the timescales of the order
of seconds. We note that by modifying the lattice potential through introducing higher order Fourier components, we are also able to 
adjust the size of individual domains.

\section{Limitations of the method} \label{Sec_limitations}

\begin{figure}
\includegraphics[width=9.5cm]{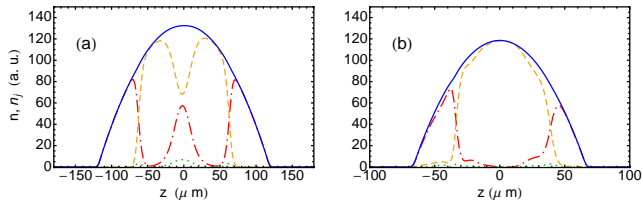}
\caption{
Same as in Fig.~\ref{fig_switchoff}(c), but for $\omega_z=2\pi\times 10\,$Hz and $N=2100$ (a) and for $\omega_z=2\pi\times 20\,$Hz and $N=1050$ (b).
}
\label{fig_limit}
\end{figure}

We now discuss some limitations of the above method of domain creation.
As we mentioned above, these metastable states are stable on an experimentally relevant time scale due to the suppression of tunneling.
However, the tunneling rate depends on the system parameters such as the trap size and atom number. In Fig.~\ref{fig_limit} we show
the result of adiabatic barrier removal for tighter harmonic traps with $\omega_z=10\,$Hz (a) and $\omega_z=20\,$Hz (b).
The atom number is here reduced proportionally to keep a similar density profile. Thus, an analogous domain structure as in
Fig.~\ref{fig_switchoff}(c) is expected.
Since the healing length, determining the size of domain boundaries (see below)
is almost the same in all cases due to a similar atom density profile, we expect the tunneling to be more significant in smaller systems.
Indeed, the effect of the tunneling is clearly visible when comparing Fig.~\ref{fig_switchoff}(c) with Fig.~\ref{fig_limit}(a) and (b).
With the decreasing distance between the domains the state becomes unstable, see Fig.~\ref{fig_limit}(a). In the case of the tightest
trap, Fig.~\ref{fig_limit}(b), the middle domain dissolves completely and the single-well ground state is obtained.

The tunneling rate can be estimated analytically along the lines of 
Ref.~\cite{Ketterle_Metastable}, where the case of a spatially varying magnetic field potential was considered.
Here, we apply a similar method to the case of a spatially varying external potential. First, we assume that the 
condensate separates into two components, one of them being $m_f=0$ and the other either $m_f=+1$ or $m_f=-1$, and neglect
the influence of the third component. The potential energy for the $m_f=i$ atoms is given by
\begin{equation}
V_i(z)=V(z) + g_i n_i + g_{ij} n_j\,,
\end{equation}
where $j$ is the other present component. The interaction constants in an $F=1$ condensate are $g_0=c_0$ and $g_{\pm 1}=g_{0\pm1}=c_0 + c_2$.
We estimate the tunneling rate of $m_f=+ 1$ atoms from the middle domain in a structure similar as the one shown in
Fig.~\ref{fig_switchoff}(c) towards the edges of the condensate. The chemical potential of atoms in the middle domain is 
$\mu_{+ 1} = V(z) + g_1 n_{+1}(z) ={\rm const}$. We assume that the density of $m_f=0$ component is negligible.
Analogously, in the neighboring domain
the $m_f=0$ atoms have chemical potential equal to $\mu_{0} = V(z) + g_0 n_{0}(z)$. The $m_f=+1$ atoms which tunnel through 
this domain feel the potential $V_{+1}(z) = V(z) + g_{01} n_{0}(z)$. Their density profile 
can be calculated in the WKB approximation 
\begin{align}
\left|\frac{d n_{+1}(z)}{dz}\right| &= -\frac{2\sqrt{2m}}{\hbar} \sqrt{V_{+1}(z) - \mu_{+1}} =    \\ \nonumber 
&=-\frac{2\sqrt{2m}}{\hbar} \sqrt{\mu_0 - \mu_{+1} + n_{0}(z)(g_{01} - g_{0})} \,,
\end{align}
where the decay of the density means that $d n_{+1}/dz >0$ on the right side of the middle domain and 
$d n_{+1}/dz <0$ on the left side. In the simplest approximation, $n_{0}(z)$ does not vary significantly and
the decay is exponential. Taking into account the pressure balance at the domain boundary, $g_0 n_0(z_0)^2 = g_1 n_{+1}(z_0)^2$,
we can estimate the decay as $n \sim \exp(-\alpha z)$ with $\alpha=2\sqrt{m c_2 n_0(z_0)}/ \hbar = 2^{(3/2)}\pi \xi_{\rm _s}^{-1}$,
where $\xi_{\rm _s}$ is the spin healing length. We can now write down the formula for the tunneling rate in the metastable state
\begin{equation} \label{dNdt}
\frac{d N}{dt} = \gamma \ee^{-\alpha \Delta z}\,,
\end{equation}
where $\Delta z$ is the distance between the domains and $\gamma$ is the
attempt rate $\gamma= 2^{(-3/2)} n_0(z_0) \langle v_{\rm s}\rangle$, with $\langle v_{\rm s}\rangle$ 
being the speed of sound averaged over the transverse density profile~\cite{Ketterle_Metastable}.
This indicates that the tunneling is greatly suppressed when the distance between the domains is larger than
the spin healing length, which determines the spin domain boundary width through the coefficient $\alpha$.
In Table~\ref{tab_limit} we show the ratio of the calculated amount of atoms that tunnel through to the total number of atoms 
in the middle domain. In the case of the tightest trap, the tunneling
over the time of the evolution would exceed the number of available atoms, which explains the absence of a metastable state.

\begin{table}[h]
\begin{tabular}{|c|c|} 
\hline 
Trap frequency & $N_{\rm tunneled} / N_{\rm total}$ \\
\hline 
$2.5\,$Hz & $10^{-12}$ \\ 
$10\,$Hz &  $0.05$ \\ 
$20\,$Hz &  $2$ \\ 
\hline 
\end{tabular} 
\caption{
The ratio of the number of atoms that can tunnel through the barrier according to Eq.~(\ref{dNdt}) within $t=5\,$s 
to the number of available atoms in the middle $m_f=+1$ domain.
}
\label{tab_limit}
\end{table}

\begin{figure}
\includegraphics[width=7.5cm]{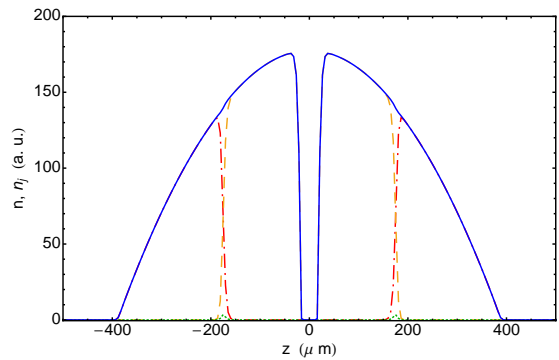}
\caption{
Ground state of the double well potential for parameters as in Fig.~\ref{fig_switchoff}(a) but with $w=9.4 \mu$m.
}
\label{fig_w1}
\end{figure}

The second limitation of the method is related to the minimal size of a domain. For illustration,
in Fig.~\ref{fig_w1} we present the condensate ground state in the double well potential as in  Fig.~\ref{fig_switchoff}(a),
but created with a very narrow Gaussian barrier. In this case the inner domains of $m_f=+1$ atoms
are absent, and consequently the slow decrease of the barrier height leads to the ground state, Fig.~\ref{fig_switchoff}(b),
instead of the metastable state, Fig.~\ref{fig_switchoff}(c). 
The reason for the absence of the domains is the very steep slope of the density profile
close to the barrier. According to Ref.~\cite{Matuszewski_Trap}, in the local density approximation (LDA), \
the spin state at a given point is determined by the value of the atom density at this point. 
The $m_f=+1$ atoms can reside only in a very narrow region on the density slope
where it is smaller than a certain value. If this region is smaller than twice the spin healing length, the domain cannot form.
The approximate size of this region can be estimated as twice the distance between the point of maximum density $n_{\rm max}$ and
half the maximum density,
\begin{equation}
\Delta z \approx \frac{2\sqrt{2} \ln 2}{\sqrt{\ln \frac{A}{c_0 n_{\rm max}}}} w\,.
\end{equation}
We confirmed through systematic numerical simulations with varying barrier size that with a good accuracy the inner domains appear
when $\Delta z$ becomes larger than $2\xi_{\rm s}$.

\section{Conclusions} \label{Sec_conclusions}

In conclusion, we demonstrated a method of controlled creation of metastable spin domains in an antiferromagnetic condensate
in a homogeneous magnetic field. The method exploits the phase separation of spin components in an external potential.
In contrast to other methods, which rely on the spatially varying magnetic fields, the domain structure is here completely
determined by optical fields, which makes this method more versatile and reconfigurable. Additionally, the method allows
for creation of domains of various sizes, with spatial resolution limited by the spin healing length only.

\acknowledgments

This work was supported by the Foundation for Polish Science through the ``Homing Plus'' programme and by the EU project NAMEQUAM.

\clearpage

\end{document}